# Controls Abstraction Towards Accelerator Physics: A Middle Layer Python Package for Particle Accelerator Control

M. King[a,b,*], A. D. Brynes[a,b], F. Jackson[a,b], J. K. Jones[a,b], N. Ziyan[a,b], M. A. Johnson[a,b], K. Baker[c], D. J. Scott[d], E. Yang[d], T. Kabana[d], C. Garnier[d], S. Chowdhury[d], N. Neveu[d], R. Roussel[d]

[a]*ASTeC, STFC, Daresbury Laboratory, Keckwick Lane, WA4 4AD, Warrington, UK*
[b]*Cockcroft Institute, Daresbury Laboratory, Keckwick Lane, WA4 4AD, Warrington, UK*
[c]*SLAC National Accelerator Laboratory, 2575 Sand Hill Rd, Menlo Park, 94025, USA*
[d]*ISIS Neutron and Muon Source, STFC, Rutherford Appleton Laboratory, Harwell Campus, OX11 0QX, Didcot, UK*

**Abstract**

Control system middle layers act as a co-ordination and communication bridge between end users, including operators, system experts, scientists, and experimental users, and the low-level control system interface. This article describes a Python package—Controls Abstraction Towards Acclerator Physics (CATAP)—which aims to build on previous experience and provide a modern Python-based middle layer with explicit abstraction, YAML-based configuration, and procedural code generation. CATAP provides a structured and coherent interface to a control system, allowing researchers and operators to centralize higher-level control logic and device information. This greatly reduces the amount of code that a user must write to perform a task, and codifies system knowledge that is usually anecdotal. The CATAP design has been deployed at two accelerator facilities, and has been developed to produce a procedurally generated facility-specific middle layer package from configuration files to enable its wider dissemination across other machines.

## 1. Introduction

Particle accelerators are highly complex machines that are comprised of hardware devices ranging from conventional systems like water, electrical, and vacuum systems to superconducting radio-frequency (RF) cavities and state-of-the-art detector systems. Many of these devices are facility-specific, non-commercial, and tailored to R&D applications. The control systems for these devices are fundamental to the operation of facilities, and their implementation can place an additional burden on the machine operator or scientist. Understanding the underlying control logic and architecture needed to perform physics-based measurements is typically not straightforward. While control systems provide the machine operator with access to the full operating range of hardware, only a small subset of hardware parameters are relevant to the operator or scientist. For example, to change the strength of a magnet in an accelerator, the scientist should not need to be aware of the various interlocks that allow the magnet power supply to be activated, or the specific mechanism for switching on the power and adjusting the current. The naming conventions used by controls engineers can also differ from those that are useful to an operator, and it can be burdensome to decipher these names in order to operate the machine. Abstracting away the complexity of adjusting and monitoring hardware, and providing methods for controlling the machine in a language that can be easily understood by operators and scientists, has been a persistent challenge.

*Corresponding author
  Email address: matthew.king@stfc.ac.uk (M. King)

Furthermore, the daily operation of these facilities requires the implementation of standard measurements and procedures, often based on scripts or high-level applications written by operators or scientists. Each of these applications must communicate with the control system to interact with hardware. If the author of each application were to work individually, a great deal of code would be (and often is) duplicated, implementations may diverge, and a change in control system could lead to multiple maintenance tasks across the codebases.

There are also additional hardware properties that may not be fully integrated or exist at all in the control system. These properties can include: the physical position and offset of an element with respect to the local coordinate system; calibration curves and factors; or 'soft' limits on parameters that may be in place temporarily. Integrating this information, some of which may be anecdotal or varying on a regular basis, into a programmatically useful and human-readable format can be a great boon for operators.

A set of tools that abstract away and translate information from the control system into a format that can be more easily understood by operators is conventionally known as a middle layer. The logic of controlling hardware can also be bundled together into simple functions, such that the developer of a script can simplify their procedures, and reduce greatly the amount of code that is required. Several middle layer frameworks have been developed previously [1], most notable of which is the Matlab Middle Layer (MML) [2] which is widely used at storage ring facilities. Additionally, a bespoke C++ middle layer was deployed during the commissioning and first experimental campaigns of the Compact Linear Accelerator for Research and Applications (CLARA) [3]. These approaches were powerful, yet limited: MML requires proprietary licenses for MATLAB, and the CLARA package lacked transferability to other facilities.

This article presents the implementation of a middle layer between the control system of a particle accelerator and the high-level applications built by operators and scientists. While the specific implementation of this design will vary between facilities, owing to differences in how the controls system and hardware have been set up, this design is generally applicable and will be of use for the wider community. The package described in this article, known as the Controls Abstraction Towards Accelerator Physics (CATAP), is a Python-based interface to the Experimental Physics and Industrial Control System (EPICS) software package [4], built on the `pyepics` library [5] for Channel Access process variables (PVs) and `p4p` [6] for PVAccess PVs [7, 8]. It has been deployed on two facilities: the Linac Coherent Light Source (LCLS) at SLAC [9], and CLARA at Daresbury Laboratory [3]. CATAP aims to build on previous experience with middle layers and to provide a modern Python-based middle layer with explicit abstraction, YAML-based configuration, and procedural code generation.

The article is structured as follows: Sec. 2 briefly outlines these two facilities, and discusses their respective software environments; Sec. 3 describes the design and architecture of the middle layer software package; Sec. 4 introduces a feature of CATAP that can create a facility-specific middle layer procedurally from YAML [10] files; Sec. 5 analyzes the development process and the deployment of the package on LCLS and CLARA; finally, Sec. 6 presents conclusions and future work.

## 2. Facility descriptions

Many commonalities can be found between hardware types in the majority of accelerator facilities, including magnets, beam position monitors, charge diagnostics, and accelerating cavities, to name only a few. However, the controls systems of the two facilities discussed in this article—LCLS and CLARA—exhibit differences, owing both to the level of development and the scale of the facilities. Therefore, this section briefly discusses these facilities and their respective control system architectures.

### 2.1. LCLS Facility Layout

The tunnel housing LCLS at SLAC contains three linear electron accelerators (linacs) that can operate simultaneously. For the remainder of this paper, the three accelerators will be referred to as LCLS-cu, LCLS-sc, and FACET-II. LCLS-cu has been operating since 2009 [9] with normal conducting traveling wave structures. LCLS-cu regularly delivers up to 16 GeV beams at a repetition rate of 120 Hz. LCLS-sc saw first light in 2023, and is ramping up to deliver 4 GeV beams at 1 MHz with superconducting cavities. During the 2026 down time, additional cavities will be added so that LCLS-sc can reach beam energies of 8 GeV. Two undulator lines (tailored for soft and hard x-rays) deliver short (fs) x-ray pulses at photon energies from about 200 eV to 2.5 keV to a variety of users (see Fig. 1 for a schematic of the LCLS facility). FACET-II is not used to generate x-rays, but does have a user community focused on advanced accelerator research, such as plasma wakefield acceleration.

Between the three linacs, there are millions of PVs associated with the RF, magnets, and diagnostics spanning a tunnel that is 3 km long. Each section of the tunnel and of the linacs in those sections are given area names. Hardware components are assigned a MAD [11] lattice name which may or may not map intuitively to the EPICS root PV names associated with that device. Here, the root PV name refers to the common start of PVs associated with a device, for example, QUAD:L1B:0385. There may be tens of PVs associated with QUAD:L1B:0385 for turning on/off the magnets, reading settings, and more, but they will all start with the same root. A strict naming convention is followed for root PV names related to common beam line elements such as magnets. MAD and base PV names are tracked in an Oracle database, which can be queried for beam line elements and a subset of metadata. The naming convention for PVs requires the area name to be included, enabling the user to determine the area of a piece of hardware based on its PV name. There are also specific names associated with common accelerator hardware, such as magnets. For example, a quadrupole magnet in the L1B area will have the root PV name of QUAD:L1B:0385, where the number 0385 is related to the location of the quadrupole with respect to other magnets and hardware in the area.

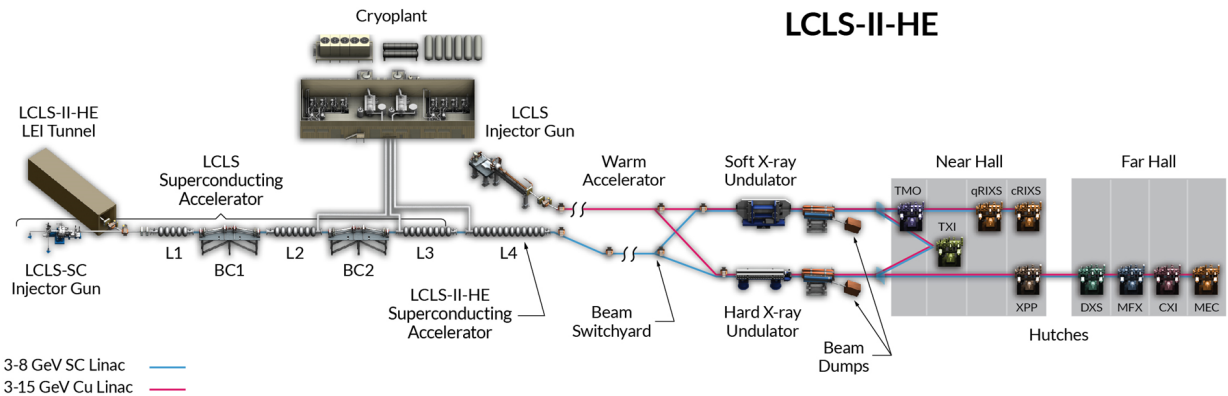

Figure 1: Schematic of the LCLS facility layout including planned High Energy (HE) upgrade. The superconducting and copper linacs can deliver beam to both the soft and hard x-ray lines.

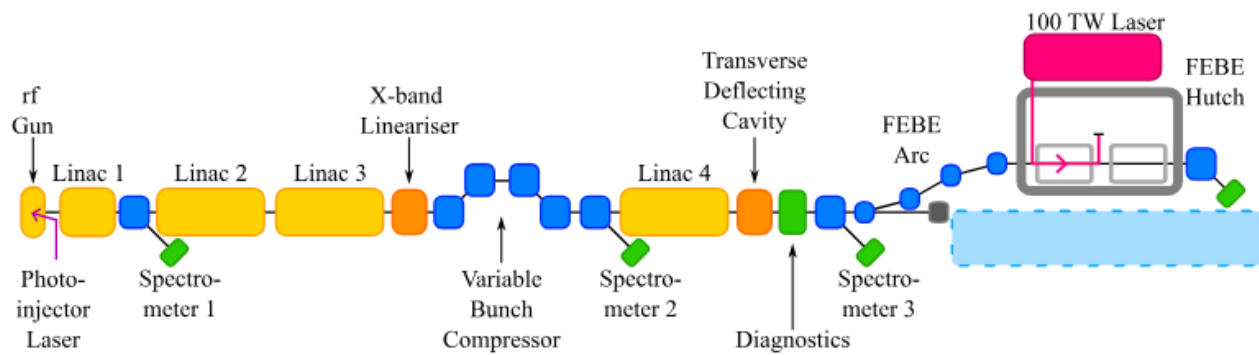

Figure 2: Schematic of the CLARA linear accelerator test facility, including the Full Energy Beam Exploitation (FEBE) beamline, shielded FEBE hutch, 100 TW laser system, and space reserved for potential future applications (shaded blue area).

2.2. CLARA Facility Layout

CLARA is a high-brightness electron beam facility based on normal-conducting RF, operating at 100 Hz and with a maximum beam energy of 250 MeV. A previous iteration of CATAP, based on C++, was deployed during commissioning of the low-energy injector [3] and the first user experiment runs. Three additional accelerating cavities and a variable magnetic bunch compressor chicane were installed in 2024, and the machine is currently undergoing commissioning before the full installation of an additional line for user experiments [12]; Fig. 2 shows the layout of the facility.

Despite being on a much smaller scale with respect to the LCLS facilities, the daily operation of CLARA requires the monitoring and control of thousands of PVs. As a facility undergoing commissioning, the control system of CLARA itself is in a state of development, with numerous new systems coming online whose control structures do not necessarily match those of previously existing hardware. Examples of this include new devices for inserting diagnostic screens, and a recently installed photoinjector laser shutter—the control of both of these hardware types has diverged from that of the low-energy injector systems commissioned in the past. Maintaining consistent naming and control interfaces across hardware and facilities that have decades of lifetime and are regularly being updated, maintained, and repaired has been a common problem when writing applications for CLARA. From the perspective of the developers of scripts, preparing the accelerator programmatically from scratch would require an awareness of all these subtle modifications, highlighting the importance of CATAP, which is able to abstract away the low-level heterogeneity to facilitate ease-of-use in the control room and consolidate any fixes or changes into a shared package.

## 3. Design and Architecture of CATAP

An accelerator lattice consists of a variety of hardware types, including magnets, accelerating cavities, and numerous categories of beam diagnostics, to name only a few. Each specific hardware object is read into the middle layer via a configuration file, which creates a Python object that has access to the control system parameters defined in that file. Similar objects are then grouped together into a higher-level object which facilitates operations on groups of hardware. The overall architecture of the middle layer is shown in Fig. 3, and the subsections below describe each aspect of the middle layer.

### 3.1. Lattice Files

The controls system information pertaining to a specific hardware object is provided to CATAP in the form of a YAML file. This human-readable configuration file is easy to version, can be used with no Python knowledge, and is independent of hardware changes. This enables CATAP and the High-Level Applications (HLA) that use it to exist outside of hardware and control system changes. Each of these files must contain two headings: `controls_information` and `properties`. The first of these contains a map of PV records—in other words, the full name of the PV, the data type (see Sec. 3.2), and a flag to indicate whether this PV is read-only. The last of these does not necessarily indicate that the PV itself is read-only, only that CATAP treats it as such. Additional further information that can be added to each PV record include the units associated with that parameter (in the case of Channel Access PVs), and a description; the former of these can be useful for calculating unit conversions, and the latter are both helpful for indexing purposes, and for producing documentation for the procedurally-generated code discussed in Sec. 4 below. It is also possible to define a protocol to use for a given PV entry, which enables communication via PVAccess or ChannelAccess. The use of PVAccess and ChannelAccess protocols can be combined within the same Hardware definition, with the interface made as similar as possible to the user. It should be noted that not every PV related to a hardware object must be defined in these files; only a subset of those parameters that are relevant should be provided.

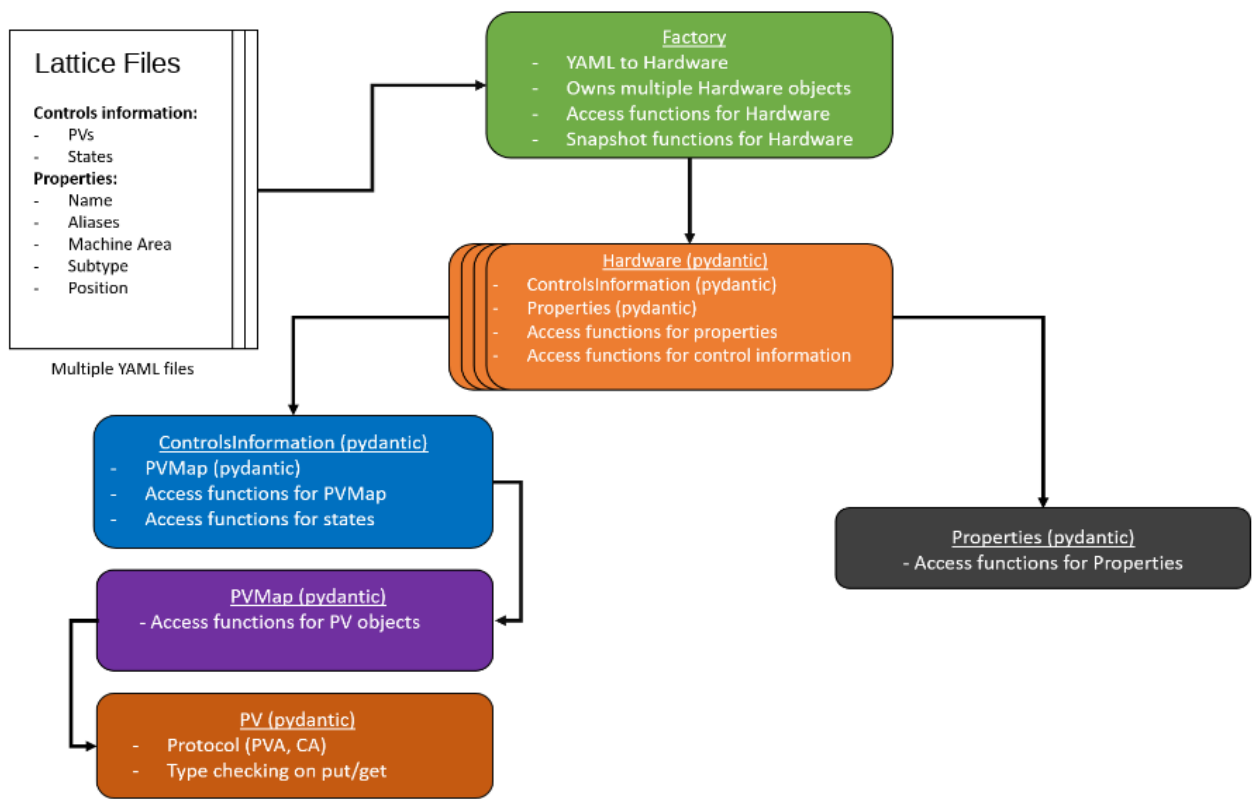

Figure 3: Base class architecture of the CATAP middle layer, showing the hierarchical structure and how a single EPICS PV is integrated into a hardware object. Stacked boxes indicate many `Hardware` instances belonging to a single `Factory`.

```yaml
# Example quadrupole magnet YAML configuration file
controls_information:
  pv_record_map:
    SETK:
      type: scalar
      read_only: False
      description: Set the strength of the quadrupole magnet.
      units: 1/m^2
      pv: <PV>:SETK
    SETI:
      [...]
    [...]
properties:
  hardware_type: Magnet
  machine_area: A1
  subtype: QUADRUPOLE
  name_alias: MAG1,QUAD1,QUADRUPOLE_1
  tolerance: 0.1
  position: 0.6
  [...]
```

Figure 4: Example YAML Configuration file for a Quadrupole magnet with Controls information and Properties defined.

The `properties` entry in the YAML file must define the following parameters: the root name of the PV, the hardware type, and the machine area and position (used for sorting). These parameters ensure that the correct hardware object type is instantiated, and they allow the higher-level classes of the middle layer to access the object. Additional custom properties to the file can be added as required. An example outline of a file to create a `Magnet` object type is provided in Fig. 4.

3.2. PV Map

Based on the PVs defined in the hardware-specific YAML file described in Sec. 3.1, connections are set up via the `pyepics` and/or `p4p` packages. Additional functionality is added over the base-level version of `pyepics` and/or `p4p`, including type-checking and validation of parameters using the `pydantic` [13] Python library. If any information is missing in the YAML file, or if a specific parameter is mis-configured, an error will be thrown. Connections to the PVs can be checked on instantiation, with a user warning being created if any issues arise. Once successfully instantiated, the properties of the `PVMap` consist of the PV objects defined in the configuration file.

The specific PV types defined in CATAP are: binary, scalar, waveform, string, state, and statistical. When using EPICS on the command line, it may be possible to write an arbitrary number to a PV that can only interpret values of 0 or 1; the middle layer will throw an exception if this type of operation is attempted. State-type PVs in the middle layer are analogous to the `enum` type. To query or write to a state PV, either an `int` or a `str` type is actionable, given that the string maps to an enum provided for that PV in the configuration file. The statistical-type PV uses the callback functionality in `pyepics`/`p4p`, meaning that its value updates automatically without the user having to query it. A buffer is also generated for this type, meaning that statistics can be calculated.

*3.3. Controls Information*

Once the `PVMap` is instantiated, a class is created containing the controls information for that hardware object. Access to data from the controls system through the `PVMap` is available at this level by creating access functions or properties for those PVs. Given that these access functions can have arbitrary names, this creates a layer of abstraction above the base-level EPICS interface, such that the user can access this data with more human-readable functionality. This includes the translation of `enum` parameters that are read back via `pyepics` as integers, such that the user can understand the readback from the control system.

Furthermore, given that similar hardware types may have different implementations in the controls system, this class is able to abstract away these differences to provide a unified method for interacting with them. Certain magnets on CLARA, for example, have a two-stage interlock chain, both of which must be cleared before the power supply can be switched on, while other magnets have only one set of interlocks. This class is able to determine the method required for clearing the interlocks, and provide a simple common interface for powering on the magnet. Similarly, there are three different movable stage types on CLARA that house diagnostic screens, along with slits and apertures, with more types envisaged for the future beamline. Depending on the specific stage type, this class is able to set the required position for the screen via a common function, regardless of the differences in the implementation in the controls system.

*3.4. Properties*

Metadata associated with a given `Hardware` (see Sec. 3.5) object are also defined in the configuration file. The name, machine area, hardware type, and position are necessary attributes of any such object, although the user can define additional properties that are useful when combined with the controls information, as well as properties that may not be recorded in a way that is accessible, for example calibration dates and installation dates. Providing the subtype of an element, such as defining a `Quadrupole` as a subset of a `Magnet` type, can enable coordinated control over similar object types. As mentioned above, certain similar hardware types may have different low-level control logic. Providing the subtype of the element, and defining appropriately this logic in the middle layer, can provide transparency for the script developer to control these various objects in a similar way. Tolerances between readbacks and setpoints can also be defined here, such that the user is made aware when the noise from a given PV is beyond an acceptable level. Another example of a useful property is a timeout, which can be used to throw an exception if a given operation does not complete in the time expected, thereby alerting the user that a given component may be faulty.

### 3.5. Hardware

The `Hardware` class incorporates both the `ControlsInformation` and the `Properties` that were instantiated based on the configuration file. This is the level at which the user of CATAP can interact with specific hardware objects on the machine. Along with the functionality provided at the lower levels, simplified and human-readable functions are provided that access the control system through its sub-classes.

More complicated logic for hardware control should be implemented at this level if required. One such example is a movable stage with a horizontal and vertical arm. In the low-energy section of CLARA there are multiple such stages, with the vertical arm housing a diagnostic screen and slits, and the horizontal arm containing apertures. If a horizontal device is in use and the user wishes to insert the screen on the vertical stage, five separate actions must be applied in the controls system, along with the monitoring of the movement states of the stage. This action, including the checking and validation of the current state of the movable stage, can instead be performed with one function in the `Screen` class (a child class of `Hardware`). A simpler example from LCLS is logically checking when actuated screens (i.e. diagnostic screens) or wires are inserted. During regular beam tuning, several screens are inserted and removed for measurement along the beam line. An automatic check to ensure that all upstream devices are retracted can be included at the hardware control level.

Another example from CLARA is the saving of an image from a camera. Before executing the `save` function in the control system, the file directory must be created, the camera acquisition state must be set, along with the number of images to acquire and the file type. After the acquisition of the image(s), the acquisition state and the saved filename must also be queried. This functionality is grouped together at the `Hardware` level for a `Camera` object into a simple `save` function.

### 3.6. Factory

The `Factory` class consists of groupings of `Hardware` objects into a single class. The configuration files for all of the hardware objects defined in the middle layer are stored in sub-directories of the main Lattice directory, grouped by hardware type. Upon instantiation, the `Factory` object creates all hardware objects of a specific type (or a subset, specified by machine area). From the `Factory`, the user can access hardware objects, using either their full name or user-friendly aliases (see Sec. 3.1), or perform operations on multiple objects simultaneously. Operations such as switching on all magnets, removing all diagnostic screens, or acquiring statistics from all beam position monitors, are thereby greatly simplified, without the user requiring specific knowledge either of all PV names or the intricacies of controlling the hardware at a low-level.

```yaml
# High-level system example
components:
  mirrors:
    - <mirror-name-1>
    - <mirror-name-2>
  shutters:
    - <shutter-name-1>
    - <shutter-name-2>
  cameras:
    - <camera-name-1>
  [...]
properties:
  hardware_type: PhotoinjectorLaser
  machine_area: LAS
  name_alias: PILaser, pilaser, PI_LASER
  virtual_cathode_camera: <camera-name-1>
  [...]
```

Figure 5: Example YAML Configuration file for a Photoinjector Laser `HighLevelSystem` with Controls information and Components defined.

### 3.7. High-Level System

Certain elements in the accelerator lattice are considered by the operator to be a single piece of hardware, whereas in the control system they exist as a group of separate hardware types. The control of these elements therefore requires performing co-ordinated operation and monitoring of disparate information. For this reason, CATAP groups together such elements into a `HighLevelSystem`. The attributes of this object consist of `Components` and `Properties`, with the former of these attributes referencing specific `Hardware` objects defined in the CATAP configuration files. These sub-components can be accessed individually via the `HighLevelSystem`, and group operations on multiple hardware types can also be performed from this level. The `HighLevelSystem` configuration can be specified, again, with a YAML lattice file (an example is provided below). Each name provided in the `components` section will be used to create the relevant hardware type using the YAML lattice files mentioned in Sec. 3.1.

The CLARA photoinjector laser exists in the middle layer as a `HighLevelSystem`, incorporating controls of the laser attenuator, energy meter, shutters, mirrors, and a virtual cathode camera, with additional manipulation of the transverse and longitudinal shaping of the laser planned for the future; see Fig. 5 for an example configuration file. Moving the transverse position of the laser on the cathode involves ensuring the shutters are open, moving the mirrors, and monitoring the laser spot position on the virtual cathode camera. These co-ordinated operations are handled much more easily by grouping `Hardware` objects in this way.

The variable magnetic bunch compressor chicane in CLARA has also been implemented as a `HighLevelSystem`. When the dipole magnets in the chicane are activated to bend the electron bunch at a given angle, the beam pipe in the chicane must move from the straight beam path on a translatable stage simultaneously. Its `Properties` contain calibration factors between the stage and the dipole strengths to facilitate this operation. Beam position monitors can be used for alignment, with trim coil magnets available for ensuring a straight trajectory. A diagnostic screen with a camera can also be used for beam measurements, and a collimator is also available. As with the photoinjector laser, harmonized simultaneous operation of these diverse hardware objects is greatly simplified through the use of a `HighLevelSystem`.

*3.8. Snapshot (Load/Save Current Machine State)*

Using all of the information stored by a `Hardware` object, or a group of objects stored in a `Factory`, all of the relevant controls system data can be captured using a `Snapshot` interface which is defined for all factories. Once again, the output format for snapshots is YAML, meaning they can be easily ingested programmatically, or opened for simple viewing by users with no programming experience. This can be used both for saving the state of a `Hardware` object, or all objects of a particular type, and for applying saved settings on to the machine. The parameters which are to be saved, and those which should be applied to the object, are specific to each object type, and are defined at the `Hardware` level. In addition to acquiring the instantaneous values of these PVs, the data buffers for statistical-type parameters (see Sec. 3.2) are also saved to the `Snapshot` object. Therefore, in addition to saving machine settings which are to be applied later, the `Snapshot` can function as a general-purpose data acquisition tool.

The user can acquire snapshots simultaneously from diverse `Hardware` or `Factory` objects and group them together in a file to capture the full state of the accelerator for measurements and data analysis, in a transparent and simple format. Functions are also provided to compare the settings between two different snapshot files, or between a file and the current machine snapshot. This enables the operator to determine which settings have changed, making it simpler to identify points of divergence between setups.

The functionality for applying the settings saved in a `Snapshot` file to the accelerator depends on the logic required for applying settings to a particular hardware object, which can be included in the specific `apply_snapshot` function. For example, it is necessary to switch on the power supply of a magnet (after clearing the interlocks, if required), and possibly first to degauss the magnet, before applying the magnetic field or strength saved in a file.

## 4. CATAP from YAML Defintions: Procedural Code Generation

Two key barriers prohibited the development of a universal code-base for earlier versions of CATAP (including the C++ version referenced in Sec. 1): there can be differences in both which parameters are used for specific hardware types (such as setting either the magnetic field for a quadrupole or the current of a power supply), and in the procedural logic for hardware (such as the preparation of a camera for saving images).

Forcing a common convention for variable and function names in CATAP would provide a unified framework for control across facilities, although this could also prevent wider adoption. An alternative approach is to generate CATAP classes procedurally based on facility-specific YAML lattice files (see Sec. 3.1), whilst maintaining the functionality of the core components. Given that LCLS and CLARA had already established naming conventions in the controls system, and that it would be unreasonable to enforce these conventions across multiple facilities, the latter option was investigated.

The architecture for the procedurally generated version of CATAP can be split up into the following sections:

1. Facility Information
2. Hardware Base Models
3. Facility Classes

Full examples of lattice files and procedurally generated code are provided in the Appendix.

*4.1. Facility Information*

The Core Classes provide the underlying structure and functionality of the middle layer that should be common between facilities, described in Secs. 3.2—3.8. The lattice/device configuration files provide the information required to generate the Base Models. The minimum set of information that must be provided for a given hardware component must contain:

- Properties: `name`, `position`, `aliases`, `machine_area`, `subtype`, `hardware type`
- Controls Information: `pv_record_map`
- PV Record Map: `description`, `type`, `pv`

*4.2. Hardware Base Models*

To generate the base models, facility-specific information pertaining to a given hardware object is provided in a YAML file and injected into a template file using the Jinja [14] Python library. The `PVMap`, `ControlsInformation`, `Properties`, `Hardware`, and `Factory BaseModel` classes are then generated for each hardware type.

The generated `BaseModel` classes contain basic implementations for getting and setting control system variables at numerous levels. This provides a method for operators to interact with the control system in a more intuitive way, with the following additional features on top of the basic `pyepics` and `p4p` functionality:

- PV Types, including Statistical PVs (with buffering);
- State mapping for `enum` types;

- Get hardware by subtype (from the `Factory` class);
- Get hardware by machine area (from the `Factory` class);
- Save, Load, Compare, and Apply settings snapshots (See 3.8).

*4.3. Facility Classes*

Beyond this initial functionality, most hardware types require facility-specific procedures to be implemented, and this is often best completed by developers at the facility. Child classes of the base models are also procedurally generated, and these are expected to be further developed by CATAP users. The functions of these classes can be overridden and new functionality can be added by developers at a given facility, depending on their specific requirements for hardware control via CATAP. See Fig. 6 for an example implementation of beam position monitor control classes based on auto-generated code. This combination of base-class templates and facility-specific YAML files can greatly facilitate the wider distribution of this framework for simplifying the way in which operators interact with large-scale experimental control systems.

## 5. Development/Implementation

With the basic components of the middle layer now outlined, this section will describe the development environment used for testing CATAP, and provide some example implementations currently deployed.

*5.1. Virtual Accelerator Controls System*

A virtual EPICS-based controls system has been developed for CLARA [15, 16] utilizing ChannelAccess and PVAccess. This provides a virtual emulation of the physical controls system, enabling the prototyping of scripts for accessing the physical machine PVs (all of which are prefixed with the string `VM-` for the virtual accelerator, or VA). While it is impractical and unnecessary to create a full duplicate of the entire controls system, most of the PVs that are relevant for the daily operation of the machine are available in the VA. The VA is hosted on a local server, meaning that switching interactions between the virtual and physical accelerator can be enabled simply by modifying the system's environment variables. Alternatively, the user can create their own VA via a Docker image.

The reading and setting of parameters via the VA is identical to that of the real controls system, with the added ability to set read-only parameters in the virtual version. For some hardware types, realistic effects such as the ramping of magnet currents has also been enabled. While noise is not implemented by default on all devices, the user can simulate this by writing to PVs that would be read-only on the physical machine at a given repetition rate.

Access to a virtual duplicate of the controls system has proven to be a great benefit for software development on CLARA, as it allows the prototyping of scripts without disrupting machine operation. The ability to simulate the control of hardware and reading back from diagnostic devices on the VA means that, when testing these measurements on the physical

```python
class BPMPVMap(BPMPVMapModel):

    ...

    @property
    def x(self):
        """ Some BPMs need to start acquiring at this facility """
        if self.is_acquiring is not None:
            if not self.is_acquiring:
                self.acquire = self.acquire_states.START
        return super().x

    @property
    def y(self):
        """ Some BPMs need to start acquiring at this facility """
        if self.is_acquiring is not None:
            if not self.is_acquiring:
                self.acquire = self.acquire_states.START
        return super().y

class BPMControlsInformation(BPMControlsInformationModel):

    ...

    @property
    def x(self) -> float:
        """ Acquiring is handled by BPMPVMap, now use calibration factor """
        return self.x * self.x_calibration_factor

    @property
    def y(self) -> float:
        """ Acquiring is handled by BPMPVMap, now use calibration factor """
        return self.y * self.y_calibration_factor
```

Figure 6: An example of a facility-specific implementation for a beam position monitor (BPM) class. The x and y functions in `BPMPVMap` have been extended to check for acquisition status. In the `BPMControlsInformation` class, x and y also utilitize the calibration factors specified in the yaml file.

machine, the majority of the work has been completed. The same is true of CATAP: by testing the package on the VA rigorously, it was relatively trivial to deploy on the physical machine.

*5.2. Photoinjector Laser Characterization*

Regular procedures carried out on the CLARA photoinjector are measurements of the electron emission of the cathode, both across the area at a fixed laser pulse energy, and as a function of laser energy. The first of these scans is performed by scanning the photoinjector laser position across the cathode surface while measuring the electron beam charge emitted from the CLARA gun. The laser position is controlled via one or more mirrors with horizontal and vertical encoders, while the mean position of the laser spot relative to the camera is determined using a virtual cathode camera. At each point along the scan, these parameters, along with the laser pulse energy and the bunch charge measured by a wall current monitor, can all be acquired simultaneously and in a co-ordinated fashion using the features of the `HighLevelSystem`. Virtual cathode images are also acquired automatically using the `Snapshot` functionality, allowing more detailed post-processing of the laser spot and its position. A similar level of control over the photoinjector is required for scanning the laser pulse energy, which is controlled using a half-wave plate and monitored using a dedicated energy meter.

*5.3. Upgrading High Level Applications at LCLS*

Several legacy high level applications (HLA) at LCLS were written in MATLAB GUIDE, and have been maintained through the superconducting upgrade. New applications for the LCLS-sc upgrade have mostly been written in Python, although with a lack of widely shared code. With the sunsetting of GUIDE by Mathworks (MATLAB), serious consideration was given to updating core MATLAB applications to Python, and how to improve their modularity. As mentioned previously, many physics measurements and applications in accelerator control rooms require similar functions or building blocks of code. The CLARA (CATAP) style design was adapted for LCLS devices and is now used to inform base device classes [17] used for the development of new HLA software. Due to the difference in scale of devices, code was written to automatically generate YAML files by area. The script generates YAML files by referencing hardware maintained in the Oracle data base. Currently, magnets, screens, wires, beam position monitors, long beam loss monitors, and transverse deflecting cavities are included in the YAML files.

Although the LCLS Python HLA applications are not as mature as CATAP, this design has helped to modularize and simplify the development efforts. Having access to device control through Python will enable the development of small convenience tools that support operations. For example, turning off all magnets in an area (10's to 100's of magnets) for a long shut down requires the repetitive opening of several PYDM panels and button pushes to turn off magnets by area. Access to `Factory`-like classes can reduce magnet shutdown work to fewer than ten lines of code (see Fig. 7), and allows for flexibility in targeted area shutdowns with little programming overhead.

```python
from lcls_tools.common.devices.reader import *

areas = ['EMIT2']

for i in areas:
    magnets = create_magnet(area=i)
    magnets.turn_off()
```

Figure 7: Example CATAP-based script for switching off all magnets in a specific area of LCLS.

There are several upgrade projects under development at SLAC now given the modularity and ease of use of CATAP.

## 6. Conclusions

This article has described the implementation of a design for simplifying the control of a particle accelerator via a middle layer between the controls system and high-level applications called CATAP (Controls Abstraction Towards Accelerator Physics). The goal of CATAP is to abstract away the low-level intricacies of the controls system so that researchers are more able to focus on the physics-related aspects of experiments. CATAP bundles together hardware objects of a similar type, and provides access functions that can be easily understood by the accelerator operator or scientist, all of which can be generated via changes to YAML files which require little to no programming ability.

The deployment of this package can greatly streamline the way in which operators interact with the machine. In general, it is no longer necessary to implement similar logic for controlling hardware multiple times across the high-level software codebase, which both reduces the burden on developers to understand low-level controls features and prevents diverging implementations of procedures.

In order to make this package more readily usable by other facilities, CATAP makes use of template files to generate the middle layer code (see Sec. 4). By providing only the PVs for a generic hardware object of a given type and their descriptions, self-documenting code can be procedurally generated that fills out all of the underlying structure and access functions specific to that object type on a given accelerator. The only remaining task for the user would then be to write functions on top of these base classes, depending on the logic of their control system (see Sec. 4). The procedurally generated version of CATAP is being investigated at CLARA, LCLS, and ISIS Neutron and Muon Spallation source. CATAP enables hardware control via both EPICS Channel Access and PV Access, with the control protocol for a given PV defined in the configuration file. The ability to modify the controls protocol at this level opens the door for others to be added, such as TANGO [18] or DOOCS [19]; this upgrade is currently under investigation.

## Acknowledgments

This work used the resources at the SLAC Shared Science Data Facility (S3DF) at SLAC National Accelerator Laboratory. S3DF is a shared High-Performance Computing facility, that supports the scientific and data-intensive computing needs at SLAC. This work was funded by DOE contract Nos. DE-SC0022559. The authors acknowledge the Auralee Edelen and Dylan Kennedy at SLAC Machine Learning group for their discussions early on in the development process of CATAP for LCLS.

# Appendix A. Example YAML Lattice Files

*Appendix A.1. BPM-01.yaml*

```yaml
# In output/yaml/BPM/BPM-01.yaml
controls_information:
  pv_record_map:
    X:
      type: statistical
      auto_buffer: True
      buffer_size: 100
      units: mm
      description: "X position reading"
      pv: "BPM-01:X_RB"
    Y:
      type: statistical
      auto_buffer: True
      buffer_size: 100
      units: mm
      description: "X position reading"
      pv: "BPM-01:Y_VAL"
  x_calibration_factor: 1.0
  y_calibration_factor: 1.0023
properties:
  name: BPM-01
  hardware_type: BPM
  subtype: Standard
```

```yaml
  position: 1.023
  machine_area: BL01
  name_alias: BL-01-BPM
  manufacturer: ACME Instruments
  model: BPM-2023
  serial_number: SN123456789
  installation_date: "2023-01-15"
  last_calibration_date: "2025-06-20"
  next_calibration_due: "2027-06-20"
```

*Appendix A.2. BPM-02.yaml*

```yaml
 In output/yaml/BPM/BPM-02.yaml
controls_information:
  pv_record_map:
    X:
      type: statistical
      protocol: PVA
      auto_buffer: True
      buffer_size: 100
      units: mm
      description: "X position reading"
      pv: "BPM-02:X_RBV"
    Y:
      type: statistical
      protocol: PVA
      auto_buffer: True
      buffer_size: 100
      units: mm
      description: "X position reading"
      pv: "BPM-02:Y_READBACK"
    ACQUIRE:
      type: state
      description: "Trigger acquisition of BPM data"
      pv: "BPM-02:ACQUIRE"
      read_only: False
      states:
        START: 1
        STOP: 0
    IS_ACQUIRING:
      type: binary
      description: "Indicates if BPM is currently acquiring data"
      pv: "BPM-02:ACQUIRE_RBV"
  x_calibration_factor: 0.02
  y_calibration_factor: 0.0023
properties:
  name: BPM-02
  hardware_type: BPM
  subtype: Standard
  position: 2.023
  machine_area: BL02
  name_alias: BL-02-BPM
  manufacturer: ACME Instruments
```

```yaml
  model: BPM-2023
  serial_number: SN987654321
  installation_date: "2023-01-15"
  last_calibration_date: "2025-06-20"
  next_calibration_due: "2026-06-20"
```

## Appendix B. Generated Base Models

*Appendix B.1. BPMPVMap Model*

```python
# In output/models/bpm.py
class BPMPVMapModel(PVMap):

    X: StatisticalPV

    """X position reading"""

    Y: StatisticalPV

    """X position reading"""

    ACQUIRE: StatePV = None

    """Trigger acquisition of BPM data"""

    IS_ACQUIRING: BinaryPV = None

    """Indicates if BPM is currently acquiring data"""

    def __init__(
        self,
        is_virtual: bool,
        connect_on_creation: bool = False,
        *args,
        **kwargs,
    ):
        BPMPVMapModel.is_virtual = is_virtual
        BPMPVMapModel.connect_on_creation = connect_on_creation
        super(
            BPMPVMapModel,
            self,
        ).__init__(
            is_virtual=is_virtual,
            connect_on_creation=connect_on_creation,
            *args,
            **kwargs,
        )

    @property
```

```python
    def x(self):
        """Default Getter implementation for X"""

        return self.X.get()

    @property
    def y(self):
        """Default Getter implementation for Y"""

        return self.Y.get()

    @property
    def acquire(self):
        """Default Getter implementation for ACQUIRE"""

        if self.ACQUIRE:
            return self.ACQUIRE.get()

    @acquire.setter
    def acquire(self, value):
        """Default Setter implementation for ACQUIRE"""

        if self.ACQUIRE:
            return self.ACQUIRE.put(value)

    @property
    def is_acquiring(self):
        """Default Getter implementation for IS_ACQUIRING"""

        if self.IS_ACQUIRING:
            return self.IS_ACQUIRING.get()

    @property
    def acquire_states(self) -> EnumMeta:
        """Default Getter implementation for :attr:`BPMPVMapModel.ACQUIRE.states`."""
        if self.ACQUIRE:
            return self.ACQUIRE.states
```

*Appendix B.2. BPMControlsInformation Model*

```python
# In output/models/bpm.py

class BPMControlsInformationModel(ControlsInformation):
    """
    Class for controlling a bpm via EPICS

    Inherits from:
        :class:`~catapcore.common.machine.hardware.ControlsInformation`
    """

    pv_record_map: SerializeAsAny[BPMPVMapModel]

    x_calibration_factor: float

    y_calibration_factor: float

    """Dictionary of PVs read in from a config file (see
    ↪ :class:`~catapcore.common.machine.hardware.PVMap`)"""
    model_config = ConfigDict(
        arbitrary_types_allowed=True,
        extra="allow",
    )

    def __init__(
        self,
        is_virtual: bool,
        connect_on_creation: bool = False,
        *args,
        **kwargs,
    ):
        BPMControlsInformationModel.is_virtual = is_virtual
        BPMControlsInformationModel.connect_on_creation = connect_on_creation
        super(
            BPMControlsInformationModel,
            self,
        ).__init__(
            is_virtual=is_virtual,
            connect_on_creation=connect_on_creation,
            *args,
            **kwargs,
        )

    @field_validator("pv_record_map", mode="before")
    @classmethod
    def validate_pv_map(cls, v: Any) -> BPMPVMapModel:
        return BPMPVMapModel(
            is_virtual=cls.is_virtual,
            connect_on_creation=cls.connect_on_creation,
            **v,
```

```python
    )

@property
def x(self):
    """Default Getter implementation for :attr:`BPMPVMapModel.X`."""
    return self.pv_record_map.x

@property
def y(self):
    """Default Getter implementation for :attr:`BPMPVMapModel.Y`."""
    return self.pv_record_map.y

@property
def acquire(self):
    """Default Getter implementation for :attr:`BPMPVMapModel.ACQUIRE`."""
    return self.pv_record_map.acquire

@acquire.setter
def acquire(self, value):
    """Default Setter implementation for :attr:`BPMPVMapModel.ACQUIRE`."""
    self.pv_record_map.acquire = value

@property
def is_acquiring(self):
    """Default Getter implementation for :attr:`BPMPVMapModel.IS_ACQUIRING`."""
    return self.pv_record_map.is_acquiring

@property
def x_calibration_factor(self) -> float:
    """Default Getter implementation for x_calibration_factor."""

    return self.x_calibration_factor

@property
def y_calibration_factor(self) -> float:
    """Default Getter implementation for y_calibration_factor."""

    return self.y_calibration_factor

@property
def acquire_states(self) -> EnumMeta:
    """Default Getter implementation for :attr:`BPMPVMapModel.acquire_states`."""
    return self.pv_record_map.acquire_states
```

*Appendix B.3. BPMProperties Model*

```python
# In output/models/bpm.py

class BPMPropertiesModel(Properties):
    """
    Class for defining bpm-specific properties.

    Inherits from:
        :class:`~catapcore.common.machine.hardware.Properties`
    """

    manufacturer: str

    model: str

    serial_number: str

    installation_date: str

    last_calibration_date: str

    next_calibration_due: str

    def __init__(self, *args, **kwargs):
        super(
            BPMPropertiesModel,
            self,
        ).__init__(
            *args,
            **kwargs,
        )
```

*Appendix B.4. BPM Model*

```python
# In output/models/bpm.py

class BPMModel(Hardware):
    """
    Middle layer class for interacting with a specific bpm object.

    Inherits from:
        :class:`~catapcore.common.machine.hardware.Hardware`
    """

    controls_information: SerializeAsAny[BPMControlsInformationModel]
    """Controls information pertaining to this bpm
    (see :class:`~catapcore.common.machine.pv_utils.ControlsInformation`)"""
    properties: SerializeAsAny[BPMPropertiesModel]
    """Properties pertaining to this bpm
    (see :class:`~catapcore.common.machine.pv_utils.Properties`)"""

    def __init__(
        self,
        is_virtual: bool,
        connect_on_creation: bool = False,
        *args,
        **kwargs,
    ):
        super(
            BPMModel,
            self,
        ).__init__(
            is_virtual=is_virtual,
            connect_on_creation=connect_on_creation,
            *args,
            **kwargs,
        )
        self._snapshot_settables = []
        self._snapshot_gettables = [
            "X",
            "Y",
            "ACQUIRE",
            "IS_ACQUIRING",
        ]

    @field_validator("controls_information", mode="before")
    @classmethod
    def validate_controls_information(cls, v: Any) -> BPMControlsInformationModel:
        try:
            return BPMControlsInformationModel(
                is_virtual=cls.is_virtual,
                connect_on_creation=cls.connect_on_creation,
                **v,
            )
```

```python
        except Exception as e:
            raise ValueError(f"Failed to validate controls_information: {e}")

    @field_validator("properties", mode="before")
    @classmethod
    def validate_properties(cls, v: Any) -> BPMPropertiesModel:
        try:
            return BPMPropertiesModel(
                **v,
            )
        except Exception as e:
            raise ValueError(f"Failed to validate properties: {e}")

    @property
    def x(self):
        """Default Getter implementation for :attr:`BPMControlsInformationModel.X`."""
        return self.controls_information.x

    @property
    def y(self):
        """Default Getter implementation for :attr:`BPMControlsInformationModel.Y`."""
        return self.controls_information.y

    @property
    def acquire(self):
        """Default Getter implementation for
          :attr:`BPMControlsInformationModel.ACQUIRE`."""
        return self.controls_information.acquire

    @acquire.setter
    def acquire(self, value):
        """Default Setter implementation for
          :attr:`BPMControlsInformationModel.ACQUIRE`."""
        self.controls_information.acquire = value

    @property
    def is_acquiring(self):
        """Default Getter implementation for
          :attr:`BPMControlsInformationModel.IS_ACQUIRING`."""
        return self.controls_information.is_acquiring
```

*Appendix B.5. BPMFactory Model*

```python
# In output/models/bpm.py

class BPMFactoryModel(Factory):
    """
    Middle layer class for interacting with multiple
    :class:`catapcore.laser.components.bpm.BPM` objects.

    Inherits from:
        :class:`~catapcore.common.machine.factory.Factory`
    """

    def __init__(
        self,
        is_virtual: bool = True,
        connect_on_creation: bool = False,
        areas: Union[MachineArea, List[MachineArea]] = None,
        hardware_type: Hardware = BPMModel,
    ):
        super(BPMFactoryModel, self).__init__(
            is_virtual=is_virtual,
            hardware_type=hardware_type,
            lattice_folder="BPM",
            connect_on_creation=connect_on_creation,
            areas=areas,
        )

    def get_bpm(self, name: Union[str, List[str]] = None) -> BPMModel:
        """
        Returns the bpm object for the given name(s).

        :param name: Name(s) of the bpm.
        :type name: str or list of str

        :return: Bpm object(s).
        :rtype: :class:`bpmModel.BPM`
        or Dict[str: :class:`bpm.BPM`]
        """
        return self.get_hardware(name)

    def x(self, name: Union[str, List[str], None] = None):
        """
        Default Getter implementation for single, multiple, all values of:
        ↪   :attr:`BPMModel.X`.

        :param name: Name(s) of the bpm.
        :type name: str or list of str or None

        :return: Value(s) of the :attr:`BPMModel.X` property.
        :rtype: property value or Dict[str, property value]
        """
```

```python
        return self._get_property(name, property_=lambda bpm: bpm.x)

    def y(self, name: Union[str, List[str], None] = None):
        """
        Default Getter implementation for single, multiple, all values of:
        ↪   :attr:`BPMModel.Y`.

        :param name: Name(s) of the bpm.
        :type name: str or list of str or None

        :return: Value(s) of the :attr:`BPMModel.Y` property.
        :rtype: property value or Dict[str, property value]
        """
        return self._get_property(name, property_=lambda bpm: bpm.y)

    def acquire(self, name: Union[str, List[str], None] = None):
        """
        Default Getter implementation for single, multiple, all values of:
        ↪   :attr:`BPMModel.ACQUIRE`.

        :param name: Name(s) of the bpm.
        :type name: str or list of str or None

        :return: Value(s) of the :attr:`BPMModel.ACQUIRE` property.
        :rtype: property value or Dict[str, property value]
        """
        return self._get_property(name, property_=lambda bpm: bpm.acquire)

    def is_acquiring(self, name: Union[str, List[str], None] = None):
        """
        Default Getter implementation for single, multiple, all values of:
        ↪   :attr:`BPMModel.IS_ACQUIRING`.

        :param name: Name(s) of the bpm.
        :type name: str or list of str or None

        :return: Value(s) of the :attr:`BPMModel.IS_ACQUIRING` property.
        :rtype: property value or Dict[str, property value]
        """
        return self._get_property(name, property_=lambda bpm: bpm.is_acquiring)
```

## Appendix C. Generated Facility Component Classes

*Appendix C.1. Component Classes*

```python
# In output/hardware/bpm.py
from models.bpm import (
    BPMPVMapModel,
    BPMControlsInformationModel,
    BPMPropertiesModel,
    BPMModel,
    BPMFactoryModel,
)
from pydantic import field_validator
from typing import Any

class BPMPVMap(BPMPVMapModel):
    """
    BPM PV Map.
    This defines the PVs used by the BPM component.

    Users should add any facility specific logic to this class.
    It inherits from the BPMPVMapModel to provide a structure.
    """

    def __init__(self, *args, **kwargs):
        super(BPMPVMap, self).__init__(*args, **kwargs)
        # Initialize any additional properties or methods specific to this model

class BPMControlsInformation(BPMControlsInformationModel):
    """
    BPM Controls Information.
    This contains the PVs and properties for the BPM component.

    Users should add any facility specific logic to this class.
    It inherits from the BPMControlsInformationModel to provide a structure.
    """

    def __init__(self, *args, **kwargs):
        super(BPMControlsInformation, self).__init__(*args, **kwargs)
        # Initialize any additional properties or methods specific to this model

    @field_validator("pv_record_map", mode="before")
    @classmethod
    def validate_pv_map(cls, v: Any) -> BPMPVMap:
        """Validate the PV Map Dictionary and Convert to PV Types"""
        return BPMPVMap(
            is_virtual=cls.is_virtual,
            connect_on_creation=cls.connect_on_creation,
            **v,
```

```python
        )

class BPMProperties(BPMPropertiesModel):
    """
    BPM Properties.
    This defines the properties of the BPM component.

    Users should add any facility specific logic to this class.
    It inherits from the BPMPropertiesModel to provide a structure.
    """

    def __init__(self, *args, **kwargs):
        super(BPMProperties, self).__init__(*args, **kwargs)
        # Initialize any additional properties or methods specific to this model

class BPM(BPMModel):
    """
    BPM.
    This represents the BPM component in the hardware layer.
    It provides access to the PVs and properties defined in the
    BPMControlsInformation and BPMProperties models.

    Users should add any facility specific logic to this class.
    It inherits from the BPMModel to provide a structure.

    """

    def __init__(self, *args, **kwargs):
        super(BPM, self).__init__(*args, **kwargs)
        # Initialize any additional properties or methods specific to this model

    @field_validator("controls_information", mode="before")
    @classmethod
    def validate_controls_information(cls, v: Any) -> BPMControlsInformation:
        """Validate the Controls Information Dictionary and Convert to
        BPMControlsInformation Type"""
        try:
            return BPMControlsInformation(
                is_virtual=cls.is_virtual,
                connect_on_creation=cls.connect_on_creation,
                **v,
            )
        except Exception as e:
            raise ValueError(f"Failed to validate controls_information: {e}")

    @field_validator("properties", mode="before")
    @classmethod
    def validate_properties(cls, v: Any) -> BPMProperties:
        """Validate the Properties Dictionary and Convert to BPMProperties Type"""
```

```python
        try:
            return BPMProperties(
                **v,
            )
        except Exception as e:
            raise ValueError(f"Failed to validate properties: {e}")

class BPMFactory(BPMFactoryModel):
    """
    BPM Factory.
    This is responsible for creating instances of the BPM component.

    Users should add any facility specific logic to this class.
    It inherits from the BPMFactoryModel to provide a structure.
    """

    def __init__(self, *args, **kwargs):
        super(BPMFactory, self).__init__(
            hardware_type=BPM,
            *args,
            **kwargs,
        )
        # Initialize any additional properties or methods specific to this model
```

*Appendix C.2. Example of Facility-specific Property Overriding*

```python
class BPMPVMap(BPMPVMapModel):

    ...

    @property
    def x(self):
        """ Some BPMs need to start acquiring at this facility """
        if self.is_acquiring is not None:
            if not self.is_acquiring:
                self.acquire = self.acquire_states.START
        return super().x

    @property
    def y(self):
        """ Some BPMs need to start acquiring at this facility """
        if self.is_acquiring is not None:
            if not self.is_acquiring:
                self.acquire = self.acquire_states.START
        return super().y

class BPMControlsInformation(BPMControlsInformationModel):

    ...

    @property
    def x(self) -> float:
        """ Acquiring is handled by BPMPVMap, now use calibration factor """
        return self.x * self.x_calibration_factor

    @property
    def y(self) -> float:
        """ Acquiring is handled by BPMPVMap, now use calibration factor """
        return self.y * self.y_calibration_factor
```

# Appendix D. Example of User Script for BPMs

```python
from output.hardware.bpm import BPMFactory
import time
bpms = BPMFactory()

# Get current values for X/Y PVs
# All PVs, Acquiring, and Calibrations are handled already!
x_readings = bpms.x()
y_readings = bpms.y()
for name in bpms.names:
    print(f"{name} X: {x_readings.get(name, None)}")
    print(f"{name} Y: {y_readings.get(name, None)}")
# Access X statistics
bpm = bpms.get_bpm("BPM-01")
# Check buffer and clear ready for new data points
print("Full Buffer?: ", bpm.x_stats.is_buffer_full)
if bpm.x_stats.is_buffer_full:
    bpm.x_stats.buffer_size = 100
    bpm.x_stats.clear_buffer()
while not bpm.x_stats.is_buffer_full:
    print("waiting for {bpm.name} x buffer to fill")
    time.sleep(1.0)
# Print out common statistics for x position
print("X Mean: ", bpm.x_stats.mean)
print("X Standard Deviation: ", bpm.x_stats.stdev)
print("X Min: ", bpm.x_stats.min)
print("X Max: ", bpm.x_stats.max)
```